\documentclass{IEEEcsmag}

\usepackage[colorlinks,urlcolor=blue,linkcolor=blue,citecolor=blue]{hyperref}
\expandafter\def\expandafter\UrlBreaks\expandafter{\UrlBreaks\do\/\do\*\do\-\do\~\do\'\do\"\do\-}
\usepackage{upmath,color}
\usepackage{tabularx} 
\usepackage{multirow}

\jvol{XX}
\jnum{XX}
\paper{8}
\jmonth{Month}
\jname{Computing in Science \& Engineering}
\jtitle{Computing in Science \& Engineering}
\pubyear{2024}

\begin{document}

\sptitle{THEME ARTICLE: CONVERGED COMPUTING: A BEST-OF-BOTH WORLDS OF HPC AND CLOUD}

\title{Secure Federated Learning Across Heterogeneous Cloud and High-Performance Computing Resources - A Case Study on Federated Fine-tuning of LLaMA 2}

\author{Zilinghan Li}
\affil{University of Illinois at Urbana-Champaign, Urbana, IL, 61820, USA}

\author{Shilan He}
\affil{University of Illinois at Urbana-Champaign, Urbana, IL, 61820, USA}

\author{Pranshu Chaturvedi}
\affil{Argonne National Laboratory, Lemont, IL, 60439, USA}

\author{Volodymyr Kindratenko}
\affil{University of Illinois at Urbana-Champaign, Urbana, IL, 61820, USA}

\author{Eliu A Huerta}
\affil{Argonne National Laboratory, Lemont, IL, 60439, USA}

\author{Kibaek Kim}
\affil{Argonne National Laboratory, Lemont, IL, 60439, USA}

\author{Ravi Madduri}
\affil{Argonne National Laboratory, Lemont, IL, 60439, USA}

\markboth{CONVERGED COMPUTING: A BEST-OF-BOTH WORLDS OF HPC AND CLOUD}{CONVERGED COMPUTING: A BEST-OF-BOTH WORLDS OF HPC AND CLOUD}

\begin{abstract}\looseness-1Federated learning enables multiple data owners to collaboratively train robust machine learning models without transferring large or sensitive local datasets by only sharing the parameters of the locally trained models. In this paper, we elaborate on the design of our Advanced Privacy-Preserving Federated Learning (APPFL) framework, which streamlines end-to-end secure and reliable federated learning experiments across cloud computing facilities and high-performance computing resources by leveraging Globus Compute, a distributed function as a service platform, and Amazon Web Services. We further demonstrate the use case of APPFL in fine-tuning a LLaMA 2 7B model using several cloud resources and supercomputers.
\end{abstract}

\maketitle

\chapteri{F}ederated learning (FL) is a distributed machine learning paradigm where multiple data owners, referred to as clients, jointly train a machine learning model.$^{1-2}$ The process is orchestrated by a central server that only requires the transfer of locally trained model parameters and not the entire datasets. The server aggregates these model parameters and redistributes the updated parameters to the clients for further local training iterations. As FL does not require collecting and storing distributed client datasets together as a centralized dataset, it is becoming an increasingly promising approach to train a more robust machine learning model and alleviate the domain shift problem without compromising the privacy of local training datasets.$^3$ FL is broadly categorized into two types, cross-device FL and cross-silo FL.$^2$ Cross-device FL involves a large number of unreliable devices, such as IoT or mobile devices, with only a small subset participating in each FL training round. On the other hand, cross-silo FL only has a few reliable clients, typically institutions or organizations equipped with powerful computing resources, including high-performance computing (HPC) systems or cloud computing facilities. This paper specifically focuses on the cross-silo FL settings.

The deployment and launch of cross-silo FL experiments face several key challenges, including the establishment of trust relationships among FL clients, inherent heterogeneity of client computing resources, and tedious coordination of the collaboration efforts. First, trust is paramount in FL to avoid data or model attacks, where a client might maliciously train the model using invalid data or send corrupted model parameters to the server. Second, the computing resources of clients in a federation can vary widely in architecture, operating systems, computing power, and job scheduling systems. Third, as cross-silo FL requires the participation of all clients in each training round, indicating the need for the simultaneous start of client training jobs to avoid resource wastage, it becomes more complex to coordinate the collaboration among multiple clients. To overcome these challenges, we introduce the Advanced Privacy-Preserving Federated Learning (APPFL) framework which enables easy and streamlined setup of secure end-to-end cross-silo FL experiments. APPFL employs \textit{Globus Compute} as its primary communication backbone for the distributed training process. \textit{Globus Compute} is a distributed function-as-a-service platform that supports a wide array of target computing systems and converts each FL client's computing machine into an endpoint,$^4$ primed for local training executions. This simplifies the process of coordinating the collaborators to initiate the FL training. Moreover, \textit{Globus Compute} is integrated with the Globus authentication service,$^5$ linking each FL client with an institutional or organizational Identity and Access Management services for identification. This facilitates building trust relationships among the clients. The APPFL framework is aimed at enabling a wide array of domain experts to easily engage in creating secure federations and running FL experiments for various scientific applications.

\section{APPFL FRAMEWORK}
\begin{figure}
\label{fig:framework}
\centerline{\includegraphics[width=18.5pc]{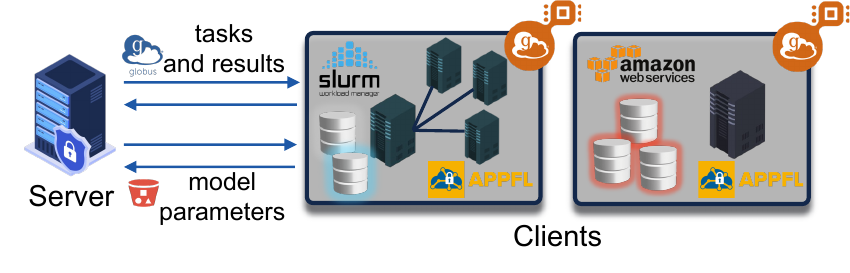}}
\caption{Overview of the federated learning process using the APPFL framework.}\vspace*{-5pt}
\end{figure}

Figure 1 illustrates the process of federated learning using the APPFL framework. In this process, the FL server plays a pivotal role, orchestrating the training by iteratively dispatching training tasks to all FL clients and subsequently gathering results via the \textit{Globus Compute} cloud server. The large model parameters are reliably exchanged via the Amazon Simple Storage Service (S3). The combination of \textit{Globus Compute}  and Amazon S3 ensures a secure, robust, and smooth flow of tasks, information, and models between the server and clients. Computing machines, ranging from personal laptops to HPC clusters with varied job schedulers, as well as cloud virtual machines, are all capable of participating as FL clients. These heterogeneous resources act as \textit{Globus Compute} endpoints to execute the dispatched training tasks using the private local datasets. Each client computing resource installs the APPFL software package, containing auxiliary codes for local training tasks. This setup highlights the versatility and adaptability of the framework to various computing environments, making it suitable for a wide range of FL applications.

Launching an FL experiment among various data owners using the APPFL framework involves a structured and secure process. The first step requires one participant to establish a Globus group, inviting other collaborating data owners through their institutional or organizational emails. This step is crucial for ensuring reliable identity and access management between the desired collaborators and FL clients, laying the foundation for an end-to-end trusted relationship. The created Globus group further provides an additional layer of authorization for FL experiments. For the actual conducting of the FL experiment, any collaborator can volunteer to take on the role of the server. This involves gathering essential information from each client, such as the \textit{Globus Compute} endpoint ID and a dataloader file. The dataloader file is particularly important as it is used for loading each client's local datasets during the remote task execution and performing the required pre-processing. Once these elements are collected, the server utilizes the training script from the APPFL framework software package to initiate the training process by providing the information collected from clients, alongside the specified model architecture and training hyperparameters. As long as all client \textit{Globus Compute} endpoints are started before the experiment initiation, \textit{Globus Compute} will automatically allocate required computing resources to execute the local training tasks, thus minimizing the complexity for coordinating the distributed training. All data owners gain access to the final model parameters at the end of the FL experiment to ensure that every participant benefits from the collaborative effort. Optionally, the experiment can be connected to the resources on Amazon Web Services that can be used to store training logs and results from various experiments along with training visualizations.

\begin{figure}
\centerline{\includegraphics[width=0.9\linewidth]{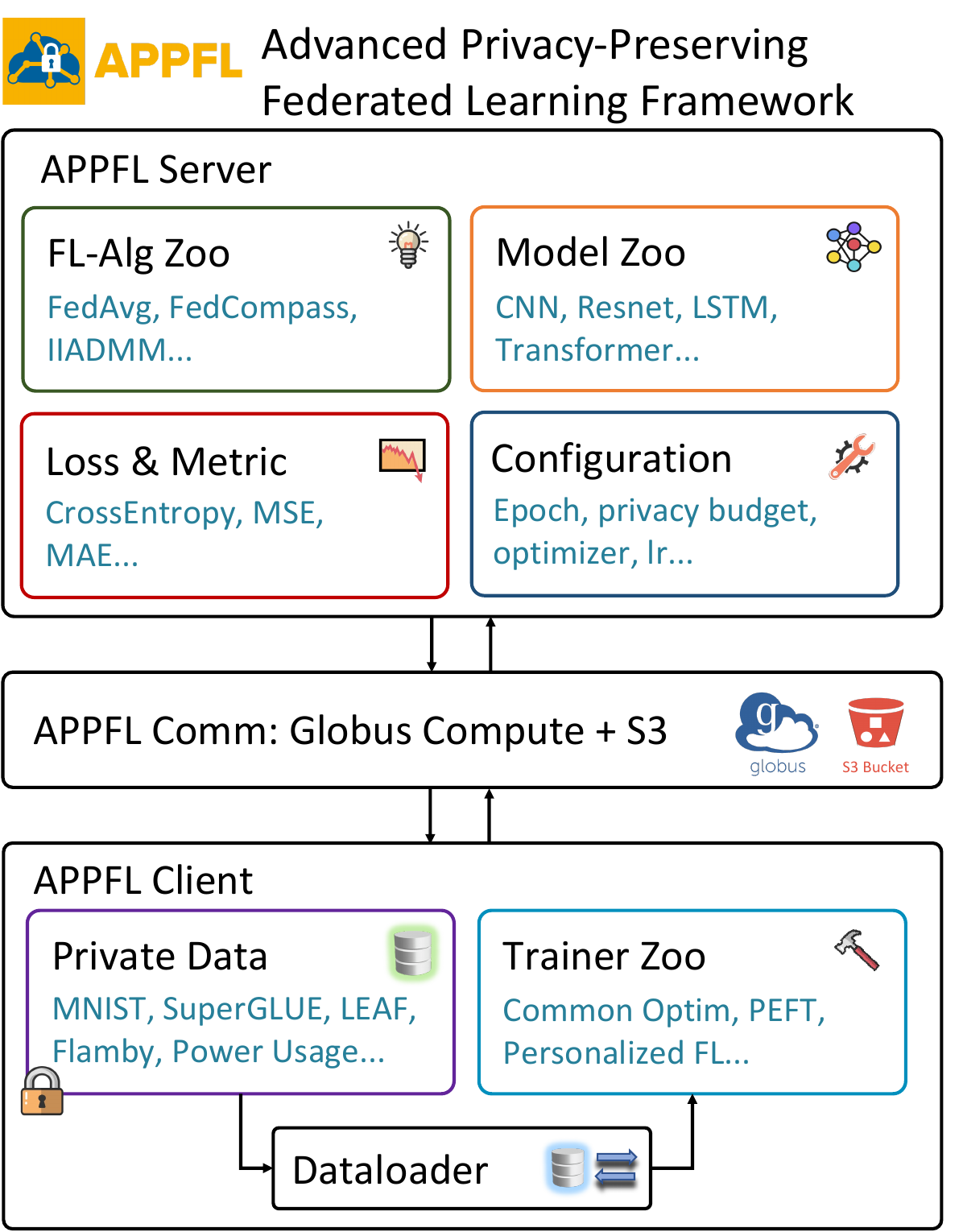}}
\caption{Modular design of the APPFL framework.}\vspace*{-5pt}
\end{figure}

Figure 2 presents the modular design of the APPFL framework. The APPFL server consists of four parts, federated learning algorithms, machine learning model architectures for training, training loss functions and metrics, and training configurations. The APPFL server supports a range of FL algorithms, including widely-used synchronous algorithms like \texttt{FedAvg},$^1$ advanced asynchronous algorithms such as \texttt{FedCompass},$^6$ and privacy-preserving algorithms such as \texttt{IIADMM}.$^7$ This versatility allows the APPFL framework to adapt to various FL scenarios and requirements. The framework incorporates several standard machine learning model architectures, including convolutional neural networks (CNN), residual neural networks (ResNet), long short-term memory networks (LSTM), and transformers. It also provides the flexibility for users to utilize custom model architectures for specific tasks. Similarly, for training loss functions and evaluation metrics, the APPFL server also offers both popular default options and the ability to use custom choices to accommodate a wide range of training scenarios. The training configuration component is for setting up necessary hyperparameters for the central aggregation and local training. On the client side, the APPFL client includes auxiliary trainers that facilitate model training on private local datasets using the provided dataloader. Multiple trainers are provided to support a variety of training tasks, from common training procedures to more specialized approaches such as parameter-efficient fine-tuning (PEFT) and personalized federated learning. The APPFL communicator lies in between the FL server and FL clients, ensuring secure and seamless communication using \textit{Globus Compute} and AWS S3 buckets.

\section{EXPERIMENTS}
To demonstrate the effectiveness of the APPFL framework in streamlining FL experiments, we present a case study focusing on the application of APPFL in federated fine-tuning the LLaMA 2 7B,$^8$ a popular open-source pre-trained large language model (LLM), on the SuperGLUE natural language understanding benchmark.$^9$ Figure 3 illustrates the overview of the experiment. Each FL client operates on an individual computing machine and accesses its local datasets. Each SuperGLUE dataset is partitioned into four client chunks in a non-independent and identically distributed manner following the \textit{dual-Dirichlet partition} strategy introduced in \texttt{FedCompass}.$^6$ The strategy employs two Dirichlet distributions to simulate the distribution of sample classes within one client (with a concentration parameter $\alpha_1=2$) and the distribution of sample sizes across clients (with a concentration parameter $\alpha_2=8$), respectively. Figure 4 illustrates how local data is distributed among the four clients for various datasets within the SuperGLUE benchmark. 

To circumvent the transfer of extensive parameters of LLM, a parameter-efficient fine-tuning (PEFT) method, low-rank adaptation (LoRA),$^{10}$ is employed. LoRA freezes all parameters of the pre-trained LLM and only trains an additional set of rank decomposition matrices injected into each transformer layer, which substantially reduces the number of trainable parameters. Specifically, for LLaMA 2 7B, with decomposition matrices applied to all query and value matrices, a rank of 8, and a scaling factor of 32, LoRA results in a total of 16.0 MB trainable parameters being exchanged between the FL server and clients. 

\begin{figure}
\centerline{\includegraphics[width=\linewidth]{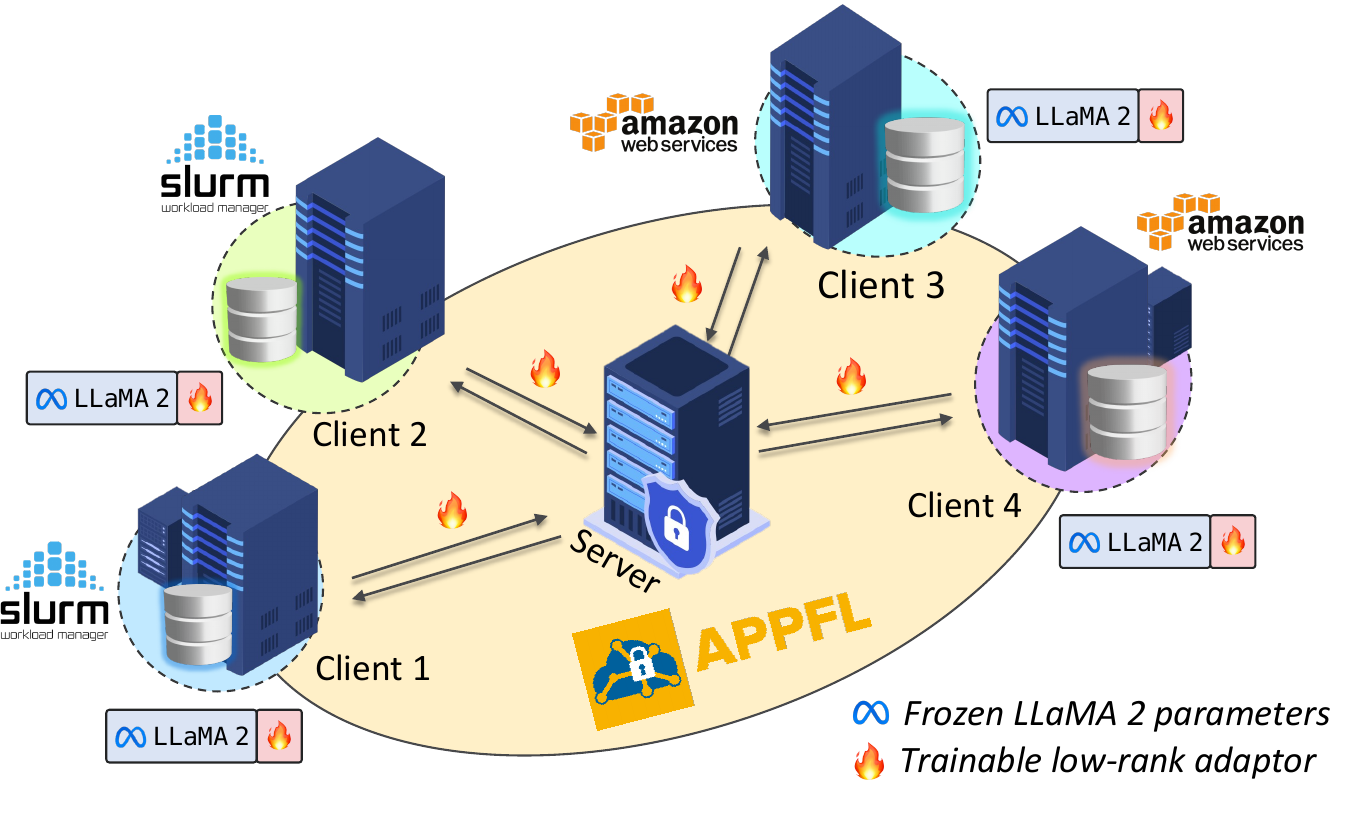}}
\caption{Overview of the federated large language model fine-tuning experiments among four heterogeneous clients on HPC nodes and cloud.}\vspace*{-5pt}
\end{figure}

\begin{figure*}[!htbp]
    \includegraphics[width=\textwidth]{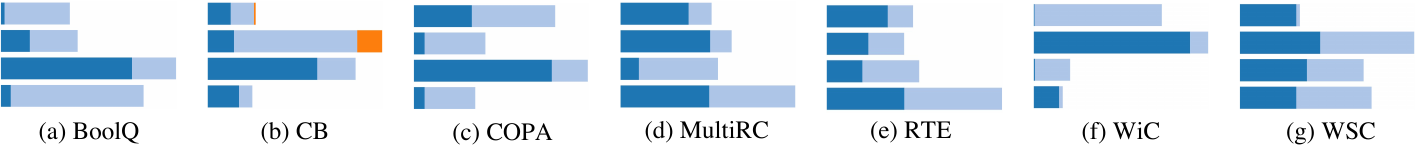}
\caption{Local data distributions among four clients for the SuperGLUE datasets, where different colors indicate samples with different labels.}
\label{fig:distribution}
\end{figure*}

\begin{table*}[t]
\centering
\caption{Detailed Stanford Alpaca prompt instructions and inputs for the SuperGLUE datasets.}
\vspace*{4pt}
\label{tab:prompts}
\begin{tabularx}{\textwidth}{|l|X|p{4cm}|}
\hline
{Dataset} & {Instruction} & {Input} \\
\hline
\small\texttt{BoolQ} & \small\texttt{The following reading comprehension question requires you to understand the following passage and answer a question related to the passage. Please answer with only "True" or "False" to the question: \{sample['question']\}?} & \small\texttt{sample['passage']} \\
\hline
\small\texttt{CB} & \small\texttt{Please determine whether the hypothesis "\{sample['hypothesis']\}" entails, contradicts, or is unrelated to the following premise: "\{sample['premise']\}". Please respond with either "Entailment", "Contradiction", or "Neutral".} & \small\texttt{N/A}\\
\hline
\small\texttt{COPA} & \small\texttt{Given the following premise, please determine whether Choice One, \{sample['choice1']\}, or Choice Two, \{sample['choice2']\}, is the \{sample['question']\} of the premise. Please respond with either "One" or "Two".} & \small\texttt{sample['premise']}\\
\hline
\small\texttt{MultiRC} & \small\texttt{Given the following paragraph, please determine whether "\{sample['answer']\}" is a correct answer to the question "\{sample['question']\}". Please respond with either "Yes" or "No".} & \small\texttt{sample['paragraph']}\\
\hline
\small\texttt{RTE} & \small\texttt{Please determine whether the sentence "\{sample['premise']\}" entails the hypothesis "\{sample['hypothesis']\}" or not. Please respond with either "Yes" or "No".} & \small\texttt{N/A}\\
\hline
\small\texttt{WiC} & \small\texttt{Please determine whether the word "\{sample['word']\}" is used in the same way in the following two sentences: "\{sample['sentence1']\}" and "\{sample['sentence2']\}" Please respond with either "Yes" or "No".} & \small\texttt{N/A}\\
\hline
\small\texttt{WSC} & \small\texttt{Please carefully read the following passages. For each passage, you must identify whether the pronoun marked in *bold* refers to the "quoted" noun.} & \small\texttt{sample['text']. \textbackslash n Question: In the passage above, does the pronoun sample['span2\_text'] refer to sample['span1\_text']}\\
\hline
\end{tabularx}
\end{table*}

\begin{table}
\centering
\caption{Number of training batches per local round and maximum token lengths for different datasets in the SuperGLUE benchmark.}
\label{tab:parames}
\vspace*{4pt}
\begin{tabular}{lcc}
\hline
{Dataset} & Batch number & {Max token length} \\
\hline
\texttt{BoolQ} & 200 & 350\\
\texttt{CB} & All & 350 \\
\texttt{COPA} & All & 300 \\
\texttt{MultiRC} & 200 & 600 \\
\texttt{RTE} & 200 & 200\\
\texttt{WiC} & 200 & 200 \\
\texttt{WSC} & All & 220 \\
\hline
\end{tabular}
\end{table}

\begin{table*}[!htbp]
\centering
\caption{Performance of the fine-tuned LLaMA 2 7B using federated learning (FL), global training, and local training.}
\label{tab:results}
\vspace*{4pt}
\begin{tabular}{lccccc}
\hline
Dataset & \# Val. Samples & \texttt{FL} (\%) & \texttt{Global} (\%) & \texttt{Local} Avg (\%) & \texttt{Local} (\%)\\
\hline
\texttt{BoolQ} & 3270 & 80.34 & 81.01 & 72.73 & [69.72, 69.45, 77.80, 73.94]\\
\texttt{CB} & 56 & 78.57 & 82.14 & 62.95 & [46.43, 76.79, 67.86, 60.71]\\
\texttt{BOPA} & 100 & 83.00 & 89.00 & 74.00 & [76.00, 68.00, 70.00, 82.00]\\
\texttt{MultiRC} & 4850 & 68.38 & 71.62 & 65.22 & [70.54, 63.72, 61.80, 64.81] \\
\texttt{RTE} & 227 & 87.36 & 85.28 & 84.66 & [86.28, 85.20, 84.48, 82.67]\\
\texttt{WiC} & 638 & 64.11 & 66.14 & 53.76 & [59.87, 50.00, 55.64, 49.53]\\
\texttt{WSC} & 104 & 72.12 & 75.96 & 64.36 & [64.42, 68.27, 57.69, 63.46]\\
\hline
\end{tabular}
\end{table*}

For all datasets in the SuperGLUE benchmark, each sample is transformed into the Stanford Alpaca prompt format.$^{11}$ Table 1 shows the detailed prompt instructions and inputs for each SuperGLUE dataset.  The APPFL PEFT local trainer minimizes the cross-entropy loss for the labeled prompt outputs using the AdamW optimizer with a learning rate of $10^{-4}$ and a decay factor of $0.85$.$^{12}$ \texttt{FedAvg} is used as the FL algorithm. The number of global communication rounds is set to 5 and the training batch size is set to 4 for all datasets. Given the varying characteristics of the datasets in the SuperGLUE benchmark, we have tailored the number of training batches of each training round and the maximum token length for each dataset in the training configurations, as detailed in Table 2. Notably, the term ``All'' for the batch number indicates that each client utilizes the entirety of available local training samples in every local training round.

To reflect real-world variability in computing resources, the four clients are operating on four heterogeneous computing machines. Specifically, two of these clients are deployed on HPC setups within the Delta supercomputer at the National Center for Supercomputing Applications (NCSA), using the Slurm job scheduler. These two differ in their GPU capabilities: one uses an NVIDIA A40 GPU, while the other employs an NVIDIA A100 Tensor Core GPU. The remaining two clients leverage Amazon Web Services (AWS) Elastic Compute Cloud (EC2) virtual machines with different specifications: one runs on a \texttt{g4ad.xlarge} instance and the other on a \texttt{g4ad.4xlarge} instance. This diverse computational setup provides a realistic testbed for the APPFL framework, demonstrating its applicability and adaptability in heterogeneous computing environments.

Table 3 presents a comparative analysis of the performance achieved by the LLaMA 2 7B model when fine-tuned under different settings: federated learning (\texttt{FL}), global training using centralized data (\texttt{Global}), and local training with client local corpus (\texttt{Local}). Since the labels for the SuperGLUE test datasets are not publicly available, the evaluation is based on the validation datasets from this benchmark. The results highlighted in the table reveal a notable pattern: models fine-tuned through FL outperform those fine-tuned locally on individual client data. This finding underscores the effectiveness of FL in enhancing model robustness. By leveraging the diverse local training corpora of various clients, FL manages to train more comprehensive models without the need for explicit data sharing. However, there remains a slight performance discrepancy when compared to models trained with centralized data, which likely arises from the inherent data heterogeneity across different clients. Additionally, the experiment showcases the versatility and capability of the APPFL framework in facilitating FL experiments across a wide range of computing environments, from HPC nodes to cloud computing facilities. This adaptability makes the APPFL framework a valuable tool for conducting FL experiments in real-world settings, where computing resources and desired training tasks can vary greatly.

\section{CONCLUSION}
In this paper, we introduce the design of the APPFL framework, a sophisticated software package to streamline the initiation and execution of secure and reliable end-to-end federated learning experiments across a diverse range of applications. This framework is adept at handling heterogeneous computing environments, from HPC systems to cloud-based resources. We showcase the framework's capabilities through a comprehensive case study, illustrating how APPFL can be seamlessly applied to the federated fine-tuning of large language models using parameter-efficient fine-tuning methods. Looking to the future, there is potential for further enhancement via improving the quality and accessibility of FL-as-a-Service provided by APPFL. Our ultimate goal is to empower a broader range of domain experts from large institutions, universities, and national laboratories, to effortlessly conduct FL experiments in various AI applications, thus expanding the horizons of collaborative, privacy-preserving AI research and development.

\section{ACKNOWLEDGMENTS}
This material is based upon work supported by the U.S. Department of Energy, Office of Science, under contract number DE-AC02-06CH11357. This research is also part of the Delta research computing project, which is supported by the National Science Foundation (award OCI 2005572), and the State of Illinois. Delta is a joint effort of the University of Illinois at Urbana-Champaign and the National Center for Supercomputing Applications.

\def\refname{REFERENCES}

\vspace*{-8pt}

\begin{IEEEbiography}{Zilinghan Li}{\,} is a Master of Science student in Computer Science at the University of Illinois at Urbana-Champaign and a research assistant at Argonne National Laboratory. His research interests include federated learning, distributed computing, and natural language processing. He received dual Bachelor's degrees from the University of Illinois at Urbana-Champaign and Zhejiang University. Contact him at zl52@illinois.edu.
\end{IEEEbiography}

\begin{IEEEbiography}{Shilan He}{\,} is a Master of Science student in Electrical and Computer Engineering at the University of Illinois at Urbana-Champaign. Her research interests include reinforcement learning, federated learning, and communication networks. She received dual Bachelor's degrees from the University of Illinois at Urbana-Champaign and Zhejiang University. Contact her at shilanh2@illinois.edu.
\end{IEEEbiography}

\begin{IEEEbiography}{Pranshu Chaturvedi}{\,} is a research engineer in the Data Science and Learning Division at Argonne National Laboratory. He is interested in trustworthy AI, federated learning, and privacy preserving machine learning. He obtained his Bachelor's degree in Computer Science and Statistics from the University of Illinois at Urbana-Champaign. Contact him at pranshu.01.c@gmail.com.
\end{IEEEbiography}

\begin{IEEEbiography}{Volodymyr Kindratenko} {\,} is an Assistant Director at the National Center for Supercomputing Applications (NCSA) at the University of Illinois at Urbana-Champaign where he serves as the Director for the Center for Artificial Intelligence Innovation (CAII). He holds an Adjunct Associate Professor appointment in the Departments of Electrical and Computer Engineering (ECE) and a Research Associate Professor appointment in the Department of Computer Science (CS). Dr. Kindratenko received a D.Sc. degree from the University of Antwerp, Belgium. His research interests include high-performance computing, special-purpose computing architectures, and machine learning systems and applications. He is an IEEE senior member. Contact him at kindrtnk@illinois.edu.
\end{IEEEbiography}

\begin{IEEEbiography}{Eliu A Huerta} {\,} is the Lead for Translational AI and Computational Scientist in the Data Science and Learning Division at Argonne National Laboratory, and CASE Senior Scientist at the Department of Computer Science at the University of Chicago. He received a Master of Advanced Study in Applied Mathematics and Theoretical Physics and a Ph.D. in Theoretical Astrophysics from the University of Cambridge, United Kingdom. His research interests are at the interface of artificial intelligence, theoretical astrophysics, extreme scale computing, scientific visualization and mathematics. Contact him at elihu@anl.gov.
\end{IEEEbiography}

\begin{IEEEbiography}{Kibaek Kim} {\,} is a Computational Mathematician in the Mathematics and Computer Science Division at Argonne National Laboratory, and a Senior Scientist at-Large at the University of Chicago Consortium for Advanced Science and Engineering. His research focuses primarily on computational optimization, including stochastic programming and integer programming, with applications to complex systems. Dr. Kim obtained a Ph.D. degree in Industrial Engineering and Management Sciences from Northwestern University. He is an IEEE senior member. Contact him at kimk@anl.gov.
\end{IEEEbiography}

\begin{IEEEbiography}{Ravi Madduri} {\,} is a Computer Scientist in the Data Science and Learning Division at Argonne National Laboratory, and a Senior Scientist at-Large at the University of Chicago Consortium for Advanced Science and Engineering. His research interests are in AI for science, large-scale computation, and biomedical informatics. He received a Master's degree in Computer Science from Illinois Institute of Technology. Contact him at madduri@anl.gov.
\end{IEEEbiography}

\end{document}